\documentclass[preprint,aps]{revtex4}
\usepackage{graphicx}

\newcommand {\bea}{\begin{eqnarray}}
\newcommand {\eea}{\end{eqnarray}}
\newcommand {\be}{\begin{equation}}
\newcommand {\ee}{\end{equation}}

\begin{document}


\title{Instantons and Monte Carlo Methods in 
Quantum Mechanics}

\author{T.~Sch\"afer$^{1,2}$}

\address{
$^1$Department of Physics, North Carolina State University,
Raleigh, NC 27695\\ 
$^2$Riken-BNL Research Center, Brookhaven National 
Laboratory, Upton, NY 11973}

\begin{abstract}
 In these lectures we describe the use of Monte Carlo
simulations in understanding the role of tunneling 
events, instantons, in a quantum mechanical toy model.
We study, in particular, a variety of methods that 
have been used in the QCD context, such as Monte Carlo
simulations of the partition function, cooling and
heating, the random and interacting instanton
liquid model, and numerical simulations of non-Gaussian
corrections to the semi-classical approximation.
 
\end{abstract}
\maketitle

\section{Introduction}
\label{sec_intro}

 We consider a non-relativistic particle moving in a potential 
$V(x)$. The Hamiltonian of the system is given by
\be
\label{H}
 H = \frac{p^2}{2m} + \lambda \left(x^2-\eta^2\right)^2 . 
\ee
We can rescale $x$ and $t$ such that $2m=\lambda=1$. We will 
also use $\hbar=1$. This means that the system is characterized 
by just one dimensionless parameter, $\eta$. The potential $V(x)$ 
with $\eta=1.4$ is shown in Fig.~\ref{fig_spec}a. The physics of 
this system is easy to understand. Classically, there are two 
degenerate minima at $x=\pm \eta$. Quantum mechanically, the two 
states can mix. If the potential barrier is very high, $\eta\to
\infty$, the wave functions of the ground state and the first 
excited state are approximately given by 
\be
\label{psipm}
\psi_{0,1}(x)=\frac{1}{\sqrt{2}}(\psi_-(x)\pm \psi_+(x)), 
\ee
where $\psi_\pm(x)$ are the ground state wave functions in the 
left and right minimum of the potential. The energy splitting
between the ground state and the first excited state is 
exponentially small. The WKB approximation gives
\be
 \Delta E = E_1-E_0 = \sqrt{\frac{6S_0}{\pi}}\omega
   \exp(-S_0),
\ee
where $\omega=4\eta$ and $S_0=m^2\omega^3/(12\lambda)=4\eta^3/3$. 

 Applications of the WKB method and of instantons to the 
double well potential are discussed in many reviews and
text books \cite{Polyakov:1976fu,Vainshtein:1981wh,Coleman:1978ae,Shuryak:1988ck,Kleinert:1995,Zinn-Justin:1993wc,Schafer:1996wv,Forkel:2000sq}.
It is not our intention to present another review on the 
subject in these lecture notes. Instead, we will use 
the double well potential to illustrate a number of 
numerical methods that have proven useful in the context
of QCD and other gauge theories. 
 
\section{Exact Diagonalization}
\label{sec_diag}

\begin{figure}
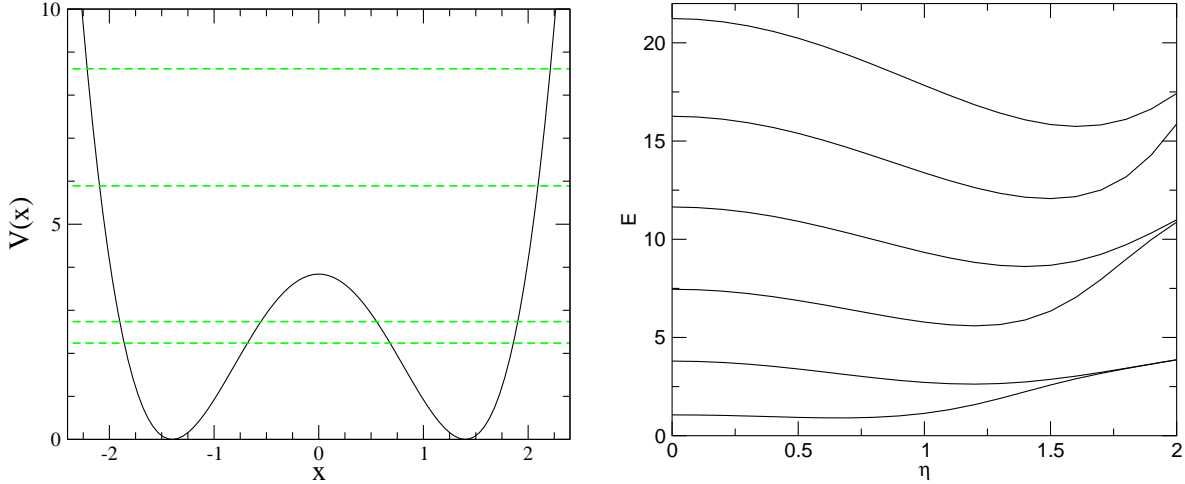

\begin{minipage}{8cm}
\includegraphics[width=7.5cm,clip=true]{pot.eps}
\end{minipage}  
\begin{minipage}{8cm}
\includegraphics[width=7.5cm,clip=true]{spectrum.eps}  
\end{minipage}
\caption{\label{fig_spec}
a) Double well potential $V(x)=(x^2-\eta^2)^2$ for $\eta=1.4$.
We have indicated the position of the ground state and the 
first three excited state. b) Spectrum of the double well potential 
as a function of the parameter $\eta$. In this figure we show 
the position of the first six states. We clearly observe that 
positive and negative parity states become degenerate as $\eta\to
\infty$.}
\end{figure}

 The quantum mechanical problem defined by the Hamiltonian 
equ.~(\ref{H}) can be solved by determining the eigenvalues 
and eigenvectors of $H$. This can be achieved by choosing a
basis and numerically diagonalizing the Hamilton operator in 
that basis. We have chosen a simple harmonic oscillator basis 
defined by the eigenstates of
\be
\label{H_0}
 H_0 = \frac{p^2}{2m}+\frac{1}{2}m\omega_0^2 x^2.
\ee
The value of $\omega_0$ is arbitrary, but the truncation 
error of the eigenvalues computed in a finite basis will 
depend on the choice of $\omega_0$. In practice, however, 
this dependence is quite weak. The eigenstates of $H_0$ 
satisfy $H_0|n\rangle = |n\rangle \omega_0(n+1/2)$. The 
Hamiltonian of the anharmonic oscillator has a very 
simple structure in this basis. The only non-zero matrix 
elements are
\bea
 \langle n|H|n\rangle &=&
    3A c^4 \left[(n+1)^2+n^2\right]
   + B c^2 (2n+1)
   + \omega_0 (n+1/2) + c ,\\
 \langle n|H|n+2\rangle &=&
     A c^4 (4n+6)\sqrt{(n+1)(n+2)}
    +B c^2 \sqrt{(n+1)(n+2)}, \\
 \langle n|H|n+4\rangle &=&
     c^4 \sqrt{(n+1)(n+2)(n+3)(n+4)},
\eea
as well as the corresponding hermitian conjugates. We have
also defined $A=1$, $B=-2\eta^2-\omega_0^2/4$, $C=\eta^4$
and $c=1/\sqrt{\omega_0}$. Note that both $H$ and $H_0$ 
conserve parity. We can decompose the matrix $H_{nm}=\langle 
n|H|m\rangle$ into even and odd components, $H=H_{even}+H_{odd}$, 
such that the eigenvectors of $H_{even}$ and $H_{odd}$ have 
positive and negative parity, respectively.

 With the choice $\omega_0=4\eta$ even modest basis sizes such 
as $N=40$ are sufficient in order to determine the first few 
eigenvectors very accurately. In Fig.~\ref{fig_spec}b we show the 
first six eigenvalues as a function of the parameter $\eta$. 
We clearly observe that as $\eta$ increases pairs of eigenvalues 
corresponding to even and odd eigenfunctions become almost 
degenerate. In this limit the eigenfunctions are of the form
given in equ.~(\ref{psipm}).

\section{Quantum Mechanics on a Euclidean Lattice }
\label{sec_latt}

 An alternative to the Hamiltonian formulation of the 
problem is the Feynman path integral \cite{Feynman}.
The path integral for the anharmonic oscillator is given by 
\be
\label{pathint}
 \langle x_1| e^{-iHt_1}|x_0\rangle = 
  \int_{x(0)=x_0}^{x(t_1)=x_1} {\cal D}x\, e^{iS}, 
  \hspace{1cm}
  S=\int_0^{t_1} dt\, \left(
   \frac{1}{4}\dot{x}^4-(x^2-\eta^2)^2 \right).
\ee
In the following we shall consider the euclidean 
partition function
\be
\label{z}
 Z(T) = \int {\cal D}x\, e^{-S_E}, \hspace{1cm}
  S_E=\int_0^{\beta} d\tau\, \left(
   \frac{1}{4}\dot{x}^4+(x^2-\eta^2)^2 \right),
\ee
where $\beta=1/T$ is the inverse temperature. The partition 
function can be expressed in terms of the eigenvalues of the 
Hamiltonian, $Z(T)=\sum_n\exp(-E_n/T)$. In the following we 
shall study numerical simulations using a discretized euclidean 
action. For this purpose we discretize the euclidean time coordinate 
$\tau_i=ia,\,i=1,\ldots n$. The discretized action is given by
\be
\label{S_disc}
 S = \sum_{i=1}^{n}\left\{ \frac{1}{4a} (x_i-x_{i-1})^2
  + a(x_i^2-\eta^2)^2 \right\},
\ee
where $x_i=x(\tau_i)$. We shall consider periodic boundary 
conditions $x_0=x_n$. The discretized euclidean path integral 
is formally equivalent to the partition function of a statistical 
system of ``spins'' $x_i$ arranged on a one-dimensional lattice. 
This statistical system can be studied using standard Monte-Carlo 
sampling methods. In the following we will simply use the Metropolis 
algorithm \cite{Creutz:1980gp,Shuryak:1984xr,Shuryak:1987tr}.
The Metropolis method generates an ensemble of configurations
$\{x_i\}^{(k)}$ where $i=1,\ldots, n$ labels the lattice points
and $k=1,\ldots,N_{conf}$ labels the configurations. Quantum 
mechanical averages are computed by averaging observables 
over many configurations, 
\be 
\langle {\cal O} \rangle = \lim_{N_{conf}\to\infty}
 \frac{1}{N_{conf}}\sum_{k=1}^{N_{conf}}
 {\cal O}^{(k)}
\ee
where ${\cal O}^{(k)}$ is the value of the classical observable
${\cal O}$ in the configuration $\{x_i\}^{(k)}$. The configurations 
are generated using Metropolis updates $\{x_i\}^{(k)}\to \{x_i\}^{(k+1)}$. 
The update consists of a sweep through the lattice during which a trial 
update $x_i^{(k+1)}= x_i^{(k)} +\delta x$ is performed for every lattice 
site. Here, $\delta x$ is a random number. The trial update is accepted 
with probability
\be
 P\left(x_i^{(k)}\to x_i^{(k+1)}\right)=
   \min\left\{\exp(-\Delta S),1\right\},
\ee 
where $\Delta S$ is the change in the action equ.~(\ref{S_disc}). 
This ensures that the configurations $\{x_i\}^{(k)}$ are distributed 
according the ``Boltzmann'' distribution $\exp(-S)$. The distribution 
of $\delta x$ is arbitrary as long as the trial update is micro-reversible, 
i.~e.~is equally likely to change $x_i^{(k)}$ to $x_i^{(k+1)}$ and back. 
The initial configuration is also arbitrary. In order to study 
equilibration it is often useful to compare an ordered (cold)
start with $\{x_i\}^{(0)}=\{\eta\}$ to a disordered (hot) start $
\{x_i\}^{(0)}=\{r_i\}$, where $r_i$ is a random variable. 

 The advantage of the Metropolis algorithm is its simplicity 
and robustness. The only parameter to adjust is the distribution
of $\delta x$. We typically take $\delta x$ to be a Gaussian random 
number with the width of the distribution adjusted such that the 
average acceptance rate for the trial updates is around $50\%$. 
Fluctuations of ${\cal O}$ provide an estimate in the error
of $\langle {\cal O}\rangle$. We have
\be 
\Delta \langle {\cal O} \rangle =
 \sqrt{\frac{\langle {\cal O}^2\rangle -\langle{\cal O}\rangle^2}
            {N_{conf}}}. 
\ee
This requires some care, because the error estimate is based on 
the assumption that the configurations are statistically independent. 
In practice this can be monitored by computing the auto-correlation 
``time'' in successive measurements ${\cal O}(\{x_i\}^{(k)})$. 
The auto-correlation time of different observables can be 
very different. For example, successive measurements of the 
total energy decorrelate very quickly, but measurements
of the topological charge have a much longer correlation
time. 

 The energy eigenvalues and wave functions of the quantum 
mechanical problem can be obtained from the euclidean correlation 
functions
\be
\label{qm_cor}
 \Pi(\tau)=\langle O(0)O(\tau)\rangle.
\ee
Here, $O(\tau)$ is an operator that can be constructed from the 
variables $x(\tau)$, e.g. some integer power $O(\tau)=
x(\tau)^n$. The euclidean correlation functions are related
to the quantum mechanical states via spectral representations.
The spectral representation is obtained by inserting a 
complete set of states into the expectation value 
equ.~(\ref{qm_cor}). We find
\be 
\label{spec}
\Pi(\tau)= \sum_n |\langle 0|O(0)|n\rangle|^2 
 \exp(-(E_n-E_0)\tau),
\ee
where $E_n$ is the energy of the state $|n\rangle$ and 
$|0\rangle$ is the ground state of the system. We can 
write this as
\be
\Pi(\tau)= \int dE\, \rho(E) \exp(-(E-E_0)\tau).
\ee
In the case we are studying here there are only bound
states and the spectral function $\rho(E)$ is a sum of 
delta-functions. Equ.~(\ref{spec}) shows that the euclidean 
correlation function is easy to construct once the energy 
eigenvalues and eigenfunctions are known. This fact was
used in order to calculate the solid lines shown in 
Figs.~(\ref{fig_cor}). The inverse problem is well defined 
in principle, but numerically much harder. In the 
following we will concentrate on extracting just the
first few energy levels. A technique that can be used 
on order determine the spectral function from euclidean
correlation functions is the maximum entropy image 
reconstruction method, see \cite{Jarrel:1996,Asakawa:2000tr}.

\begin{figure}
\begin{center}
\leavevmode
\includegraphics[width=9cm]{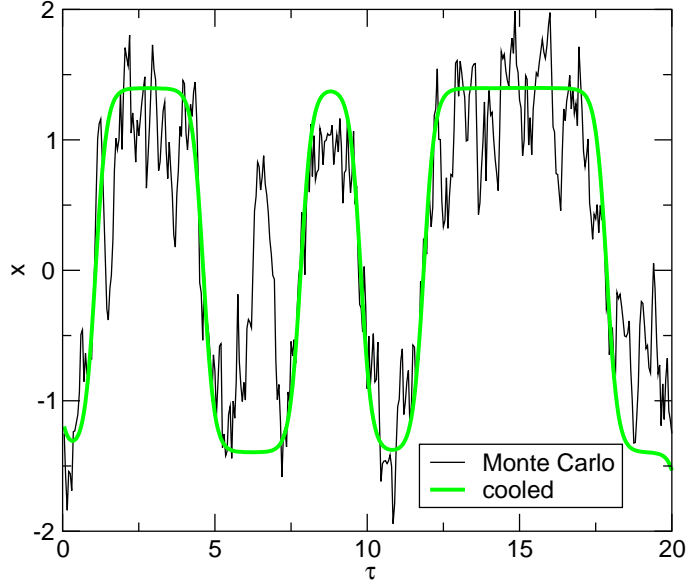}  
\end{center}  
\caption{\label{fig_path}
Typical euclidean path obtained in a Monte Carlo simulation
of the discretized euclidean action of the double well 
potential for $\eta=1.4$. The lattice spacing in the 
euclidean time direction is $a=0.05$ and the total number 
of lattice points is $N_\tau=800$. The green curve shows 
the corresponding smooth path obtained by running 100 
cooling sweeps on the original path. }
\end{figure}

 The Monte Carlo method is very useful in calculating 
expectation values in quantum or statistical mechanics. 
However, the Monte Carlo method does not directly give 
the partition function or the free energy. In principle
one can reconstruct the free energy from the energy 
eigenvalues but this is not very practical since, as 
we just mentioned, it is hard to compute the full spectrum. 
A very effective method for computing the free energy is 
the adiabatic switching technique. The idea is to start 
from a reference system for which the free energy is 
known and calculate the free energy difference to the 
real system using Monte Carlo methods. 

 For this purpose we write the action as $S_\alpha=S_0+\alpha
\Delta S$ where $S$ is the full action, $S_0$ is the action
of the reference system, $\Delta S$ is defined by $\Delta S
=S-S_0$, and $\alpha$ is a coupling constant. The action 
$S_\alpha$ interpolates between the real and the reference 
system. Integrating the relation $\partial \log Z(\alpha)/
(\partial\alpha)=-\langle \Delta S \rangle_\alpha$ we find
\be
\label{adiab}
 \log(Z(\alpha=1))=\log(Z(\alpha=0)) 
 - \int_0^1 d\alpha'\, \langle \Delta S\rangle_{\alpha'} \;\; ,
\ee
where $\langle .\rangle_\alpha$ is an expectation value 
calculated using the action $S_\alpha$. In the present case 
it is natural to use the harmonic oscillator as a reference
system. In that case
\be 
 Z(\alpha=0) = \sum_n \exp(-\beta E_n^0) 
  = \frac{\exp(-\beta\omega_0/2)}{1-\exp(-\beta\omega_0)},
\ee
where $\omega_0$ is the oscillator constant. Note that the
free energy of the anharmonic oscillator should be independent
of $\omega_0$. The integral over the coupling constant $\alpha$ 
can easily be calculated in Monte Carlo simulations by slowly
changing $\alpha$ from 0 to 1 during the simulation. In 
order to estimate systematic errors due to incomplete equilibration 
it is useful to repeat the calculation with $\alpha$ changing 
from 1 to 0 and study possible hysteresis effects. 

\section{Numerical Results}
\label{sec_num}

\begin{figure}
\begin{center}
\leavevmode
\includegraphics[width=9cm]{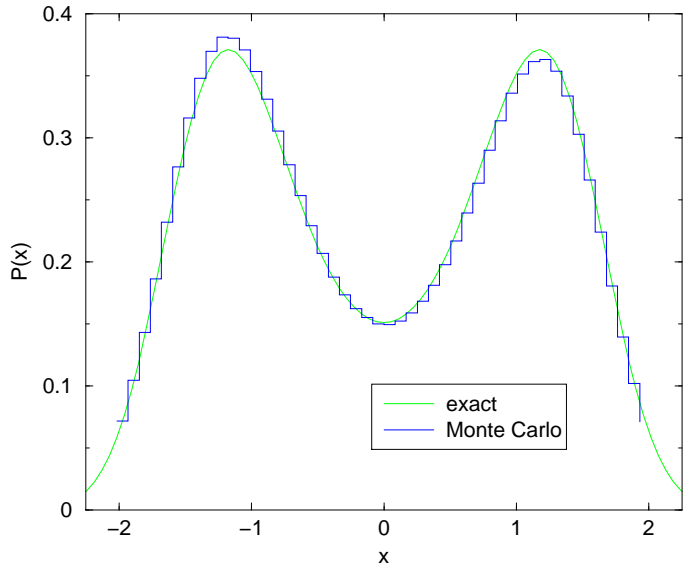}  
\end{center}  
\caption{\label{fig_psi}
Probability distribution $|\psi(x)|^2$ in the double well 
potential for $\eta=1.4$. The solid line shows the ``exact''
numerical result obtained by diagonalizing the Hamiltonian
in an oscillator basis whereas the histogram shows the 
distribution of $x$ for an ensemble of euclidean paths. }
\end{figure}

\begin{figure}
\begin{center}
\leavevmode
\includegraphics[width=8cm,clip=true]{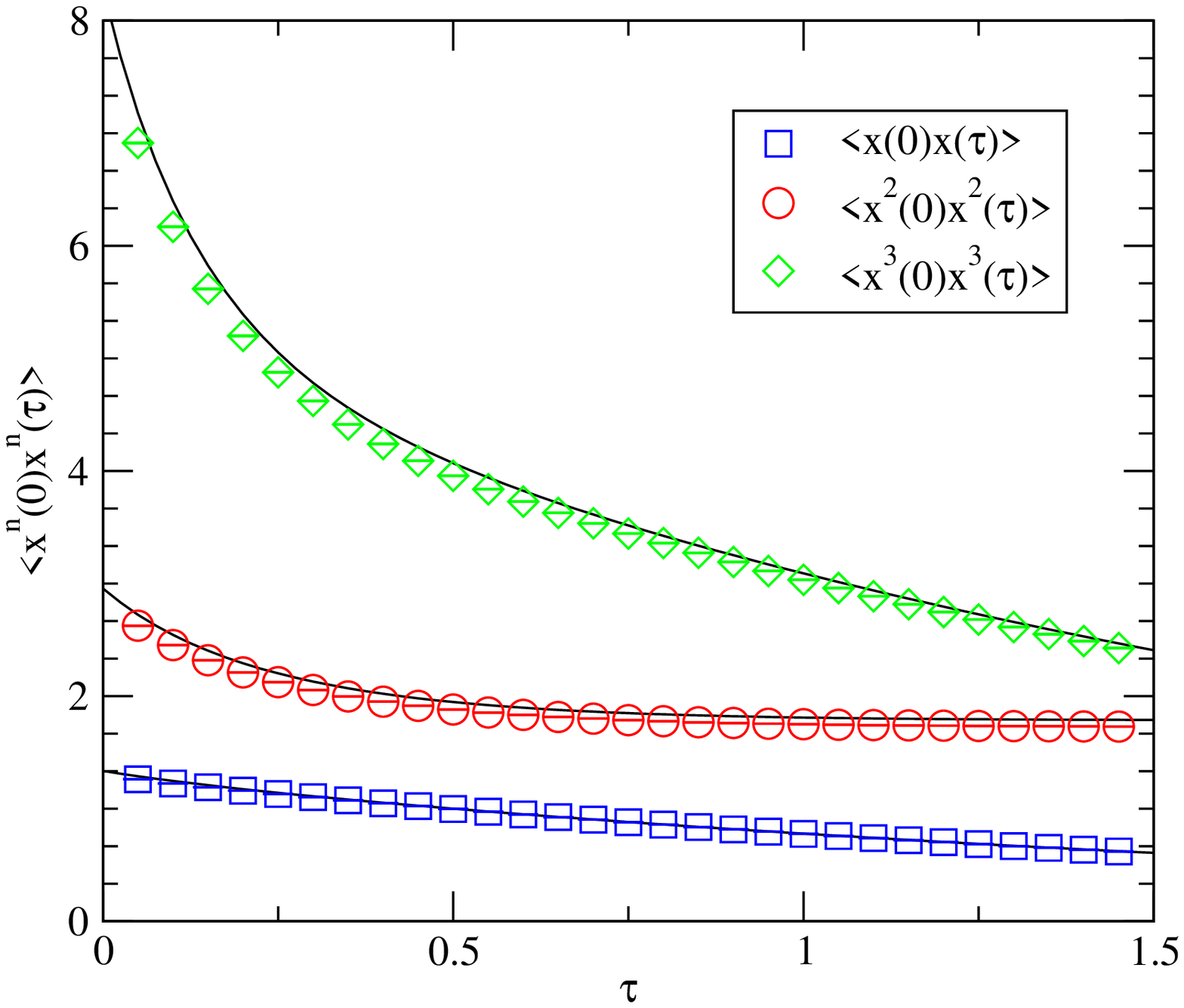}  
\includegraphics[width=8cm,clip=true]{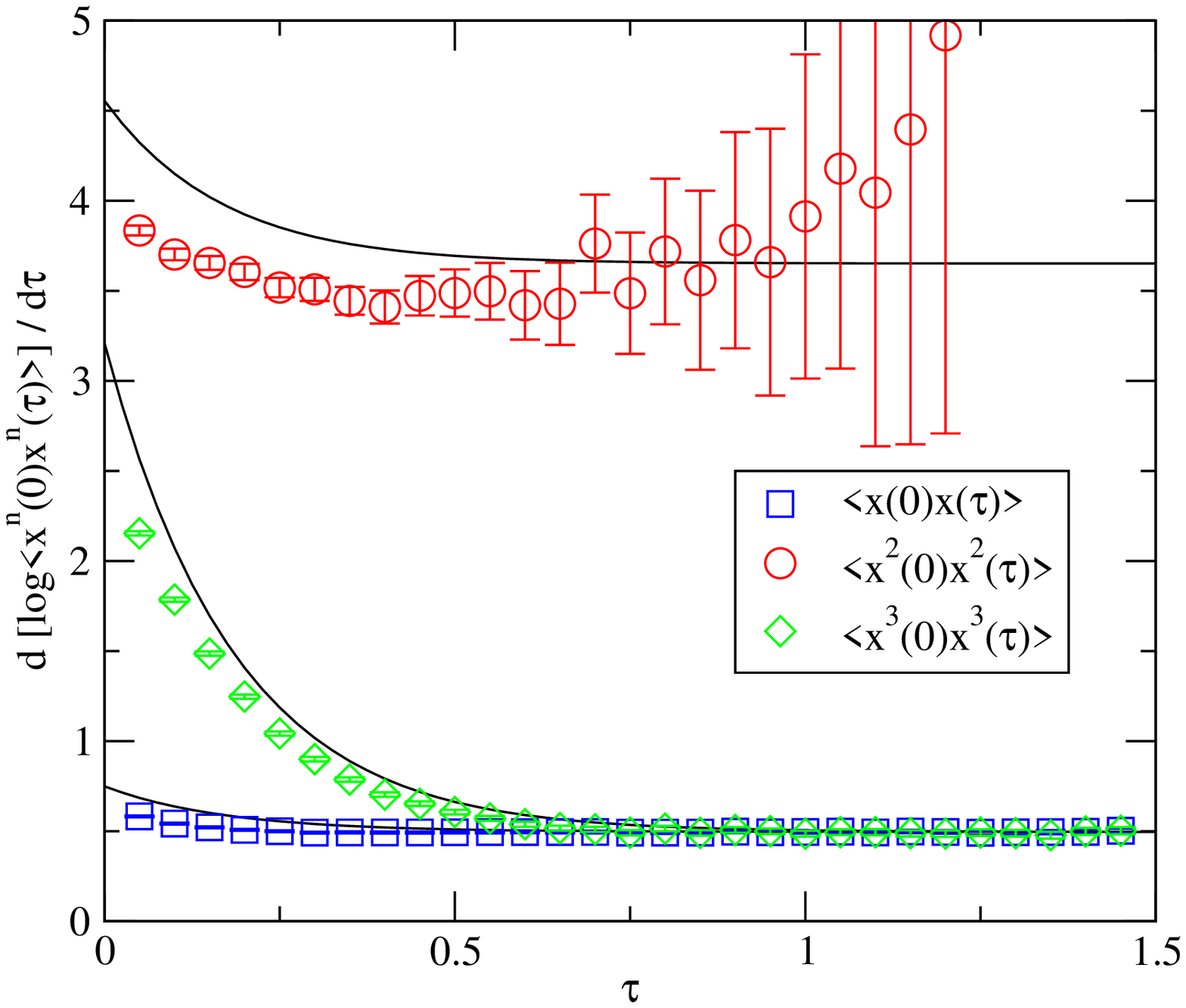}  
\end{center}  
\caption{\label{fig_cor}
Fig.~a shows the correlation functions $\langle {\cal O}(0)
{\cal O}(\tau)\rangle$ in the double well potential for 
$\eta=1.4$ and ${\cal O}=x,x^2,x^3$. The solid lines are ``exact'' 
numerical results obtained by diagonalizing the Hamiltonian in an 
oscillator basis whereas the data point were obtained from  Monte
Carlo simulations with $a=0.05$ and $N_\tau=800$. Fig.~b shows
the logarithmic derivative of the correlators in Fig.~a. In the 
case of the $\langle x^2(0)x^2(\tau)\rangle$ we subtracted the
constant contribution.} 
\end{figure}

  Numerical results from Monte Carlo simulations of the euclidean 
path integral are shown in Figs.~\ref{fig_path}-\ref{fig_cor}. The 
numerical data were obtained using the program {\tt qm.for} which 
is described in more detail in the appendix. A typical path that 
appears in the Monte Carlo simulation is shown in Fig.~\ref{fig_path}. 
The figure clearly shows that there are two characteristic time scales 
in the problem. On short time scales the motion is controlled by the
oscillation time $\tau_{osc}\sim \omega^{-1}\sim (4\eta)^{-1}$. 
For large $\tau$ the system is governed by the tunneling time 
$\tau_{tun} \sim \exp(-4\eta^3/3)$. In order to perform reliable
simulations we have to make sure that the lattice spacing $a$ is 
small compared to $\tau_{osc}$ and that the total length of 
the lattice $Na$ is much larger than the tunneling time
\be
 a\ll \tau_{osc}, \hspace{0.5cm} \tau_{tun}\ll Na .
\ee
A typical choice of parameters for the case $\eta=1.4$ is a number 
of lattice points $N=800$, a lattice spacing $a=0.05$ and a number 
of Metropolis sweeps $N_{conf}=10^5$. 

  Fig.~\ref{fig_psi} shows the distribution of $x_i$ obtained 
in the Monte Carlo simulation compared to the square of the 
ground state wave function computed by the diagonalization 
method discussed in section \ref{sec_diag}. As $\eta$ is 
increased and the potential barrier becomes larger the tunneling
time increases exponentially and the number of configurations
needed to reproduce the correct wave function becomes very
large. 

 Fig.~\ref{fig_cor} shows the correlation functions of the 
operators $x,x^2$ and $x^3$. The solid lines show the result 
obtained using the spectral representation equ.~(\ref{spec})
together with the eigenvalues and eigenfunctions determined
by numerical diagonalization of the Hamiltonian. The data 
points show the results from the Monte Carlo simulation. There 
is a small systematic disagreement for small $\tau$ which is 
related to discretization errors but the overall agreement 
is excellent. Energy levels and matrix elements can be obtained
from the logarithmic derivative of the correlation function, 
\be
 C(\tau)= -\frac{d\log\Pi(\tau)}{d\tau} = 
 \frac{\sum_n (E_n-E_0) |\langle 0|O(0)|n\rangle|^2 
 \exp(-(E_n-E_0)\tau)}
      {\sum_n |\langle 0|O(0)|n\rangle|^2 \exp(-(E_n-E_0)\tau)}.
\ee
In the limit $\tau\to\infty$ the function $C(\tau)$ converges to 
the energy splitting between the ground state and the first excited 
state that has a non-vanishing transition amplitude $\langle 0|O(0)
|n\rangle$. Because of parity invariance, $O=x^n$ connects the 
groundstate to parity even/odd levels for $n$ even/odd. Since the 
first excited state is parity odd we have
\be 
\lim_{\tau\to\infty} \frac{d}{d\tau} \log \langle x(\tau)x(0)\rangle 
 = \lim_{\tau\to\infty}\frac{d}{d\tau} \log \langle x^3(\tau)x^3(0)\rangle 
  = E_1-E_0.
\ee
For even powers of $x$ the situation is more complicated
because the correlator has a constant term $|\langle 0|x^{2n}
|0\rangle|^2$. After subtracting the constant part, the logarithmic 
derivative of the correlation function of even powers of $x$ tends 
to $(E_2-E_0)$. Numerical results are shown in Fig.~\ref{fig_cor}b. 
We observe that the logarithmic derivative of $\langle x(\tau) x(0)
\rangle$ converges very rapidly to $\Delta E_1=(E_1-E_0)$. The 
numerical results for $\Delta E_2=(E_2-E_0)$ have large 
uncertainties. These uncertainties are related to the fact that 
the correlator $\langle x^2(\tau)x(0)\rangle$ is dominated by
the subtraction constant $\langle x^2\rangle^2$. The logarithmic
derivative of the $\langle x^3(\tau)x^3(0)\rangle$ correlator
also tends to $\Delta E_1$, but receives larger contributions 
from excited states. This feature can be used in order to extract 
the energies of higher states. The idea is very simple. From the 
matrix elements $c_1=\langle 0|x|1\rangle$ and $d_1=\langle 0|
x^3|1\rangle$ we can determine a new operator ${\cal O}=x/c_1-
x^3/d_1$ that does not couple to the first excited state. This 
operator predominantly couples to the third excited state. 
Repeating this procedure we can determine the energies of 
higher excited states. The problem is that correlation
functions of higher powers of $x$ are more and more noisy.
As a result, finding the energies of highly excited states
is very hard, even in the simple problem considered here.

\begin{figure}
\begin{center}
\includegraphics[width=9cm,clip=true]{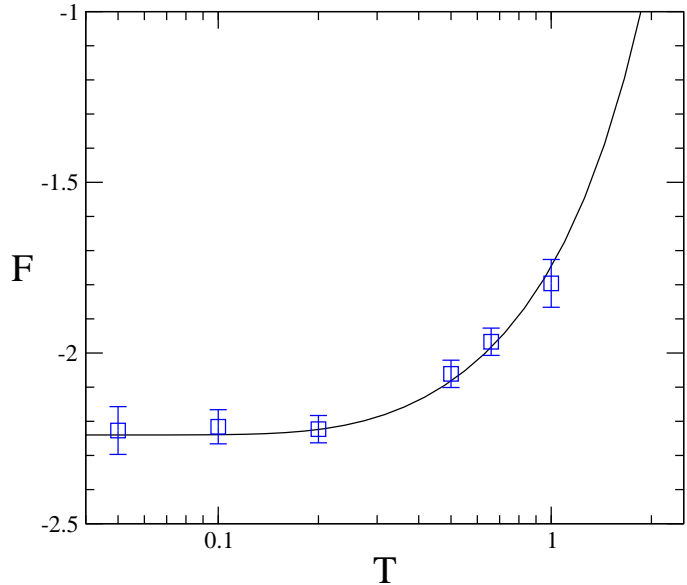}  
\end{center}  
\caption{\label{fig_F}
Free energy $F=-T\log(Z)$ of the anharmonic oscillator 
as a function of the temperature $T=1/\beta$ with $\beta
=na$. The solid line was calculated using the spectrum 
of the Hamiltonian. The data points were obtained using 
Monte Carlo calculations and the adiabatic switching 
method.}
\end{figure}

 In Fig.~\ref{fig_F} we show Monte Carlo results for the 
partition function compared to ``exact'' results based on 
the spectrum of the anharmonic oscillator obtained in 
Sect.~\ref{sec_diag}. The Monte Carlo results agree with 
the direct calculations but the Monte Carlo method is 
effectively limited to a small range of temperatures.
If the temperature is very small the partition function
is dominated by the ground state contribution. In that case,
it is much more efficient to compute the ground state 
energy directly by measuring the expectation value of the 
Hamiltonian, $E_0=\langle H\rangle$ , with $H=\dot{x}^2/4
+V(x)$. There is one subtlety with this approach: If a 
naive one-sided discretization of the time derivative is 
used then the continuum limit of the expectation value 
of the kinetic energy diverges. This problem can be 
addressed by using an improved discretization of the kinetic
energy \cite{Feynman}, or by using the Virial theorem.
The Virial theorem implies that 
\be 
\langle H \rangle = \langle T+V \rangle
 = \langle \frac{x}{2}V'+V \rangle .
\ee
At high temperature more and more states contribute. The 
main difficulty with the Monte Carlo approach in this 
regime is that discretization errors have to be carefully 
monitored.

\section{Extracting the Instanton Content using Cooling}
\label{sec_cool}

 From Fig.~\ref{fig_path} we can clearly see that for this particular
choice of the parameter $\eta$ a typical path contains two 
components, one related to quantum fluctuations with frequency
$\omega$, and one related to tunneling events, instantons. In 
the continuum limit the instanton solution can be found from 
the classical equation of motion
\be 
\frac{\delta}{\delta x(\tau)} S_E = 0
\hspace{0.5cm}\Rightarrow\hspace{0.5cm}
m\ddot{x}=V'(x).
\ee
\begin{figure}
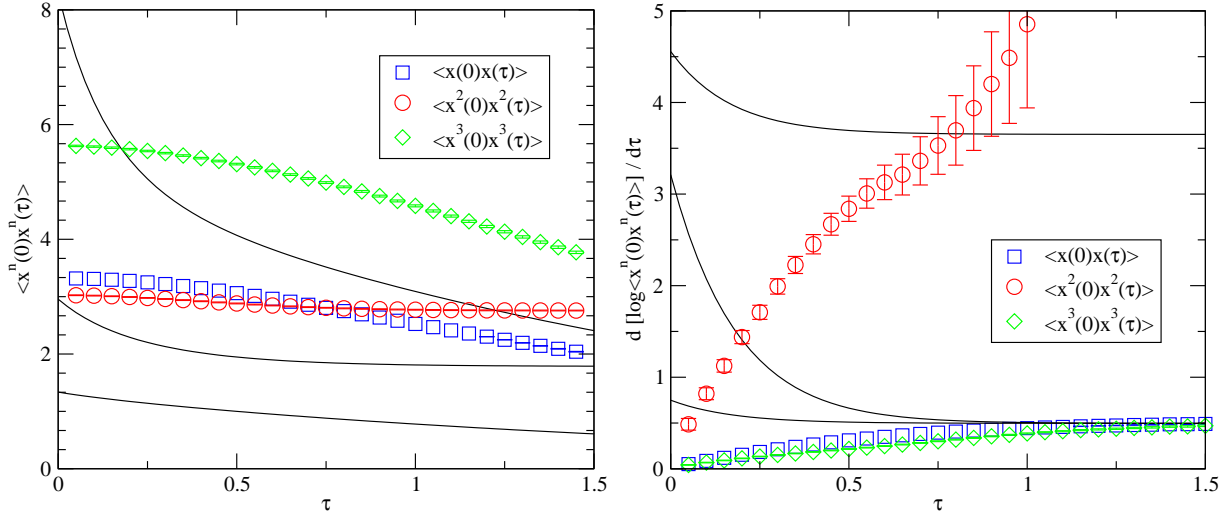

\begin{center}
\leavevmode
\includegraphics[width=8cm,clip=true]{qmcoolcor.eps}  
\includegraphics[width=8cm,clip=true]{dqmcoolcor.eps}  
\end{center}  
\caption{\label{fig_cor_cool}
Same as Fig.~\ref{fig_cor} but the correlation functions are 
evaluated from cooled Monte Carlo configurations. The 
number of cooling sweeps is $N_{cool}=200$. }
\end{figure}
The solution which satisfies the boundary condition 
$x(\tau\to\pm\infty)=\pm\eta$ is given by
\be 
x_{I}(\tau) = \eta\tanh\left[
  \frac{\omega}{2}(\tau-\tau_0)\right],
\ee
where $\omega=4\eta$ and $\tau_0$ is the ``location'' of the 
instanton. The anti-instanton solution is simply given by 
$x_{A}(\tau)=-x_I(\tau)$. The classical action of the instanton is 
\be
 S_0 = \frac{4\eta^3}{3} .
\ee
The tunneling rate $n_{I+A}=N_{I+A}/\beta$ is exponentially
small, $n_{I+A}\sim\exp(-S_0)$. In order to determine the 
pre-factor one has to study small fluctuations around the 
instanton solution. This calculation has been carried out 
to next-to-leading order in the semi-classical expansion. The 
result is \cite{Zinn-Justin:1993wc,Kleinert:1995,Wohler:pg}
\be 
\label{idens}
n_{I+A}= 8\eta^{5/2} \sqrt{\frac{2}{\pi}}
 \exp\left(-S_0-\frac{71}{72}\frac{1}{S_0}\right).
\ee
The tunneling events can be studied in more detail after removing
short distance fluctuations. A well known method for doing this
is ``cooling'' \cite{Hoek:1985gj,Hoek:1986hq}. In the cooling method 
we only accept Metropolis updates that lower the action. This will 
drive the system towards the nearest classical solution. Since instantons 
are classical solutions, cooling can be used to study the instanton
content of a quantum configurations. This is clearly seen 
in Fig.~\ref{fig_path}. The black line is the original, quantum,
configuration. The green line is the same configuration after
200 cooling sweeps. It is easy to check that this configuration
is very close to a linear superposition of independent 
tunneling events. For this purpose we can extract the instanton
and anti-instanton locations from the zero crossings and 
compare the cooled configuration to the simple ``sum ansatz''
\be 
\label{sum}
x_{sum}(\tau) =  \eta \left\{ \sum_{i} Q_i\tanh\left[
  \frac{\omega}{2}(\tau-\tau_i)\right] -1 \right\},
\ee
where $Q_i=\pm 1$ is the topological charge of the instanton. 
The most important question is to what extent physical observables
in the cooled configurations resemble those in the original 
configurations. This provides a measure of the importance of 
instantons in the double well potential. In Fig.~\ref{fig_cor_cool}
we show correlation functions measured in the cooled configurations. 
These results should be compared with the full correlation functions 
shown in Fig.~\ref{fig_cor}. We observe that the correlation functions 
are quite different. Short distance fluctuations eliminated by cooling 
obviously play an important role. We observe, however, that the level
splitting between the ground state and the first excited 
state is clearly dominated by semi-classical configurations. 
The logarithmic derivative of both the $\langle x(0)x(\tau)
\rangle$ and $\langle x^3(0)x^3(\tau)\rangle$ correlation
functions is very well reproduced in the cooled configurations.

\section{The density of instantons}
\label{sec_dens}

 The cooling method can also be used in order to get an 
estimate of the total density of instantons and anti-instantons. 
While the net topological charge, the number of instantons
minus the number of anti-instantons, is unambiguously defined
the same is not true for the total number of topological objects. 
There is no clear distinction between a large quantum fluctuation
and a very close instanton-anti-instanton pair. In the cooling 
method this is reflected by the fact that the number of instantons, 
extracted from the number of zero crossings in the cooled 
configuration, depends on the number of cooling sweeps. 

\begin{figure}
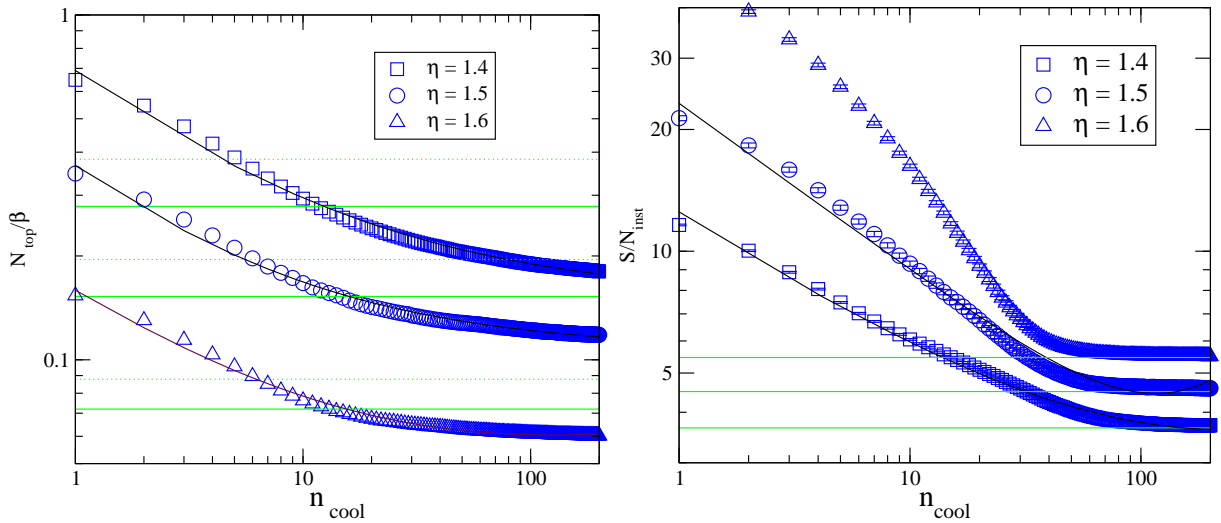

\begin{center}
\leavevmode
\includegraphics[width=8cm,clip=true]{idens.eps}  
\includegraphics[width=8cm,clip=true]{sinst_f.eps}  
\end{center}  
\caption{\label{fig_n_cool}
Instanton density and instanton action as a function of the 
number of cooling sweeps for different values of the parameter 
$\eta$. The solid and dashed green lines in Fig.~a shows the 
semi-classical instanton density at one and two-loop order. 
The solid line in Fig.~b shows the classical instanton action.}
\end{figure}

 It is clear, however, that the instanton density should be
well defined in the semi-classical limit. In this limit there is an 
exponentially large separation of scales between the tunneling 
time $\tau_{tun}$ and the scale of ordinary quantum fluctuations
$\tau_{osc}$. This separation of scales can also be exploited 
in the cooling method. Cooling is a local algorithm which implies
that it takes on the order of $\tau/a$ cooling sweeps in order 
to affect coherent structures that exist at a scale $\tau$. We
expect that the number of instantons measured using the cooling
method is approximately given by the sum of two exponentials, 
$N_I(n_{cool})=N_{osc}\exp(-n_{cool}a/\tau_{osc})+N_{tun}\exp(-n_{cool}
a/\tau_{tun})$. The first exponential describes the disappearance
of quantum fluctuations on a time scale $\tau_{osc}$ and the 
second exponential reflects instanton-anti-instanton annihilation
occurring on a time scale $\tau_{tun}$. 

\begin{figure}
\begin{center}
\leavevmode
\includegraphics[width=9cm,clip=true]{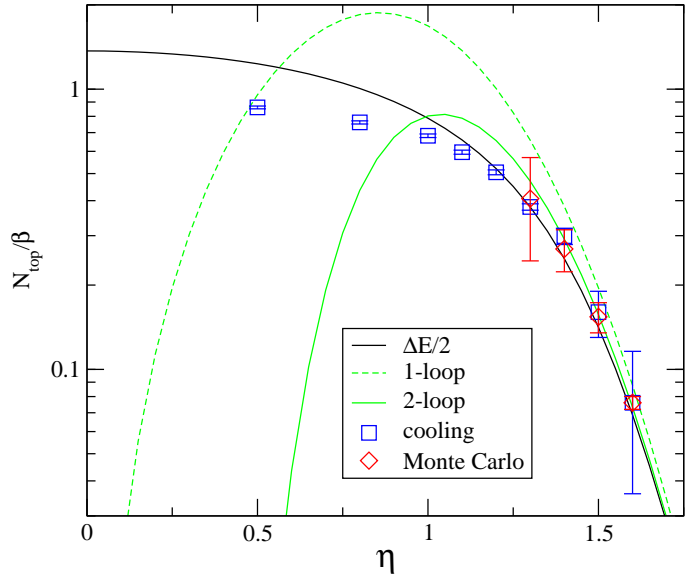}  
\end{center}  
\caption{\label{fig_idens}
Instanton density as a function of the parameter $\eta$.
The blue symbols show the instanton density extracted 
from Monte Carlo configurations after 10 cooling sweeps. 
The red symbols show the results of a Monte Carlo calculation
of non-Gaussian effects. The green lines show the semi-classical 
instanton density at one and two-loop order. The black line shows 
$\Delta E/2$ where $\Delta E$ is the splitting between the ground 
state and the first excited state.}
\end{figure}

 Numerical results for $N_I(n_{cool})$ are shown in Fig.~\ref{fig_n_cool}. 
We observe that the data is consistent with the presence of two 
distinct time scales and that the description in terms of two 
exponentials becomes better as the semi-classical limit $\eta\to
\infty$ is approached. We also note that after the quantum noise
has disappeared the instanton density is close to the semi-classical
result equ.~(\ref{idens}). A more detailed comparison is shown 
in Fig.~\ref{fig_idens}. In this figure we show the instanton
density after 10 cooling sweeps, the one and two-loop semi-classical 
result, as well as the level spacing between the ground state and
the first excited state. 

 We observe that for $\eta>1.2$, corresponding to a classical 
instanton action $S_0>2$, the number of instantons extracted 
using the cooling method agrees very well with the semi-classical
approximation. We also note that the two-loop result is a 
clear improvement over the one-loop approximation for classical 
actions as small as $S_0\sim 1$. Finally, we observe that the 
instanton density is close to the level splitting even in the
regime where $S_0$ is less than one. 

\begin{figure}
\begin{center}
\leavevmode
\includegraphics[width=9cm,clip=true]{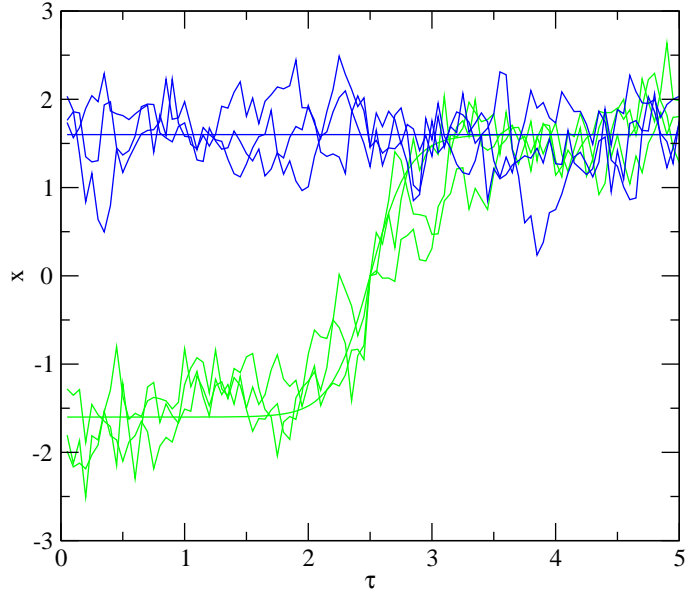}    
\end{center}  
\caption{\label{fig_ngauss}
Quantum mechanical paths which appear in a Monte-Carlo 
calculation of the one-instanton partition function 
in the double well potential. The calculation involves
adiabatic switching between the Gaussian effective 
potential and the full potential. The smooth curves
are the initial configurations in the zero and one-instanton
sector. The Monte Carlo updates in the one-instanton 
sector involve a constraint which keeps the instanton
location fixed.}
\end{figure}
  
 In Fig.~\ref{fig_n_cool} we also show a Monte Carlo calculation
of the instanton density on a small lattice. The idea is very 
simple. The one-loop calculation of the tunneling rate is 
based on expanding the action around the classical path to 
quadratic order 
\be 
\label{s_gauss}
 S = S_0 + \frac{1}{2} \int d\tau\, 
 \delta x(\tau) \left.\frac{\delta^2 S}{\delta x^2}
  \right|_{x_I(\tau)}  \delta x(\tau) + \ldots ,
\ee
where $\delta x(\tau)=x(\tau)-x_I(\tau)$. As in Sect.~\ref{sec_latt}
we can introduce a new action $S_\alpha$ that interpolates between
the full action and the Gaussian approximation, $S_\alpha= S_{gauss}
+\alpha \Delta S$ with $\Delta S= S-S_{gauss}$. The exact quantum
weight of an instanton can be determined by integrating over the
coupling constant $\alpha$. We have
\be
 n = n_{gauss}\exp\left[
  -\int_0^1 d\alpha  \left( \langle \Delta S\rangle_\alpha^{(1)}
  - \langle \Delta S\rangle_\alpha^{(0)} \right)\right],
\ee
where $\langle .\rangle_\alpha^{(n)}$ is an expectation value
in the $n$-instanton sector at coupling $\alpha$. The method 
is illustrated in Fig.~\ref{fig_ngauss}. The figure shows typical
paths that contribute to $\langle \Delta S\rangle$ in the zero
and one-instanton sector. The resulting estimate of the instanton
density is also shown in Fig.~\ref{fig_idens}. The Monte Carlo 
results show that the instanton density is reduced compared
to the one-loop estimate. For classical instanton actions 
$S_0>3$ the result is in agreement with the two-loop estimate
and the cooling calculation. It is hard to push the Monte Carlo
calculation to instanton actions $S_0<3$ because transitions
between the zero and two (four, six, $\ldots$) sector become
too frequent. 

\section{The instanton liquid model}
\label{sec_ilm}

\begin{figure}
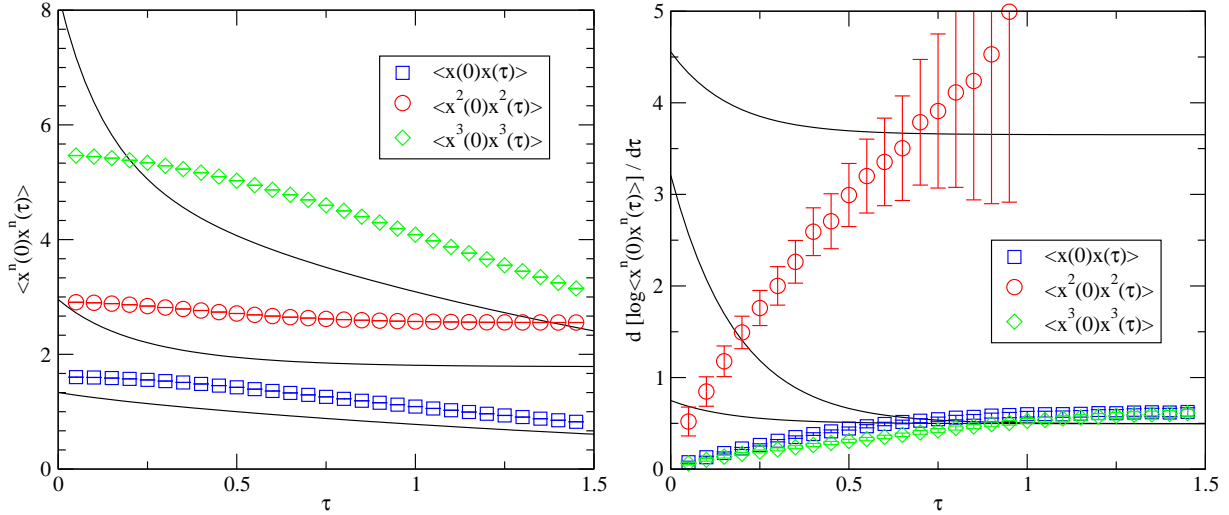

\begin{center}
\leavevmode
\includegraphics[width=8cm,clip=true]{rcor.eps}  
\includegraphics[width=8cm,clip=true]{drcor.eps}  
\end{center}  
\caption{\label{fig_cor_rilm}
Same as Fig.~\ref{fig_cor} but the correlation functions are 
evaluated from a random instanton configuration.}
\end{figure}

 Given the success of the semi-classical approximation in predicting 
the splitting between the ground state and the first excited state
it seems natural to study the correlation functions in the semi-classical
approximation in more detail. We begin by considering the contribution
from the classical path only. In this case the partition function is 
given by
\be 
\label{rilm}
Z = \sum_{n_I,n_A}\frac{\delta_{n_I,n_A}}{n_I!n_A!}\left(\prod_{i}
  \int d\tau_i\right) \exp(-S).
\ee
Here, $n_I,n_A$ are the number of instanton and anti-instantons, 
$\tau_i$ are the (anti) instanton positions, and $S_0$ is the 
classical action. In the next section we will discuss the 
problem of choosing the correct path for a multi-instanton 
configuration in more detail. The simplest choice is the sum 
ansatz given in equ.~(\ref{sum}). The coordinate correlation 
function is given by 
\be 
\Pi_{cl}(\tau) = \langle x_{cl}(0)x_{cl}(\tau) \rangle ,
\ee
where $\langle . \rangle$ denotes an ensemble average over 
the collective coordinates $\tau_i$. The distribution of 
collective coordinates is controlled by the partition function
equ.~(\ref{rilm}). The simplest approximation is to ignore 
the interaction between instantons. In this case the action 
is $S=(n_I+n_A)S_0$ and the distribution of collective 
coordinates is random. This is known as the instanton gas
model or the random instanton approximation. 

\begin{figure}
\begin{center}
\leavevmode
\includegraphics[width=9cm,clip=true]{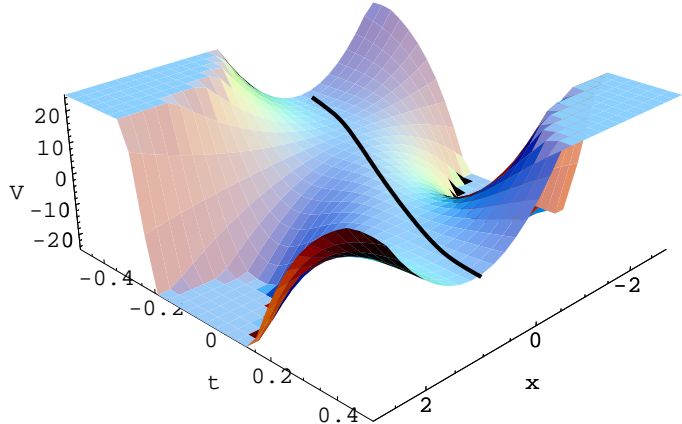}    
\end{center}  
\caption{\label{fig_veff}
Gaussian effective potential for small fluctuations
around a single instanton path centered at $\tau=0$. }
\end{figure}

 Correlation functions in a random instanton gas are shown 
in Fig.~\ref{fig_cor_rilm}. We note that the result is 
very similar to the cooled correlation functions shown 
in Fig.~\ref{fig_cor_cool}. This is in agreement with 
our earlier observation that the cooled configurations
are very close to a simple superposition of instantons. 
Like the the cooling calculation the random instanton 
gas reproduces the splitting between the ground state 
and the first excited state, but it does not give a
good description of other aspects of the correlation 
functions. 

\begin{figure}
\begin{center}
\leavevmode
\includegraphics[width=9cm,clip=true]{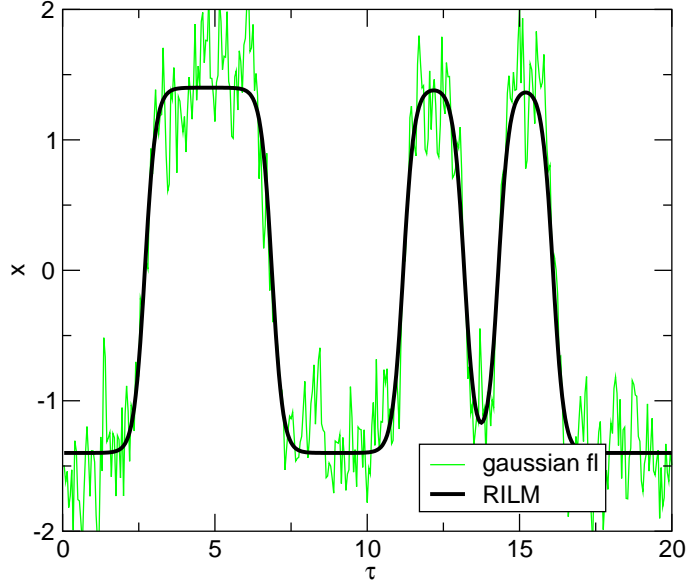}    
\end{center}  
\caption{\label{fig_conf_gauss}
Typical random instanton configuration and the same 
configuration with Gaussian fluctuations. The noisy 
path was generated using 10 heating sweeps in the 
Gaussian potential around the classical path. This figure 
should be compared with Fig.~\ref{fig_path}.}
\end{figure}

 It is clear that the main feature that is missing from the 
ensemble of classical paths is quantum fluctuations. Quantum 
fluctuations appear at next order in the semi-classical
approximation. We already noted that quantum fluctuations
determine the pre-exponential factor in the tunneling 
rate, see equ.~(\ref{idens}). We can write the path as
$x(\tau)=x_{cl}(\tau)+\delta x(\tau)$ where $x_{cl}(\tau)$
is the classical path and $\delta x(\tau)$ is the fluctuating 
part. To second order in $\delta x$ the action is given by
equ.~(\ref{s_gauss}). For a single instanton it is possible
to determine the propagator $\langle \delta x(0)\delta x(\tau)
\rangle$ analytically, see equ.~(39) in \cite{Schafer:1996wv}. 
For an ensemble of instantons we can approximate the 
propagator as a sum of contributions due to individual 
instantons. This is the procedure that is used in the 
QCD calculations described in 
\cite{Shuryak:1992jz,Shuryak:1992ke,Schafer:1995pz,Schafer:1995uz}.

\begin{figure}
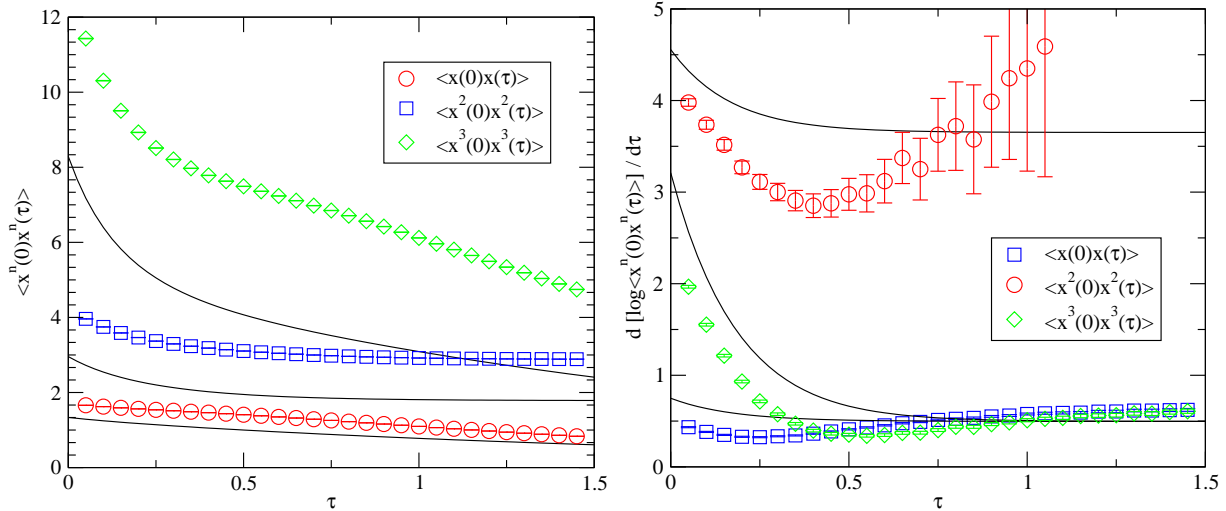

\begin{center}
\leavevmode
\includegraphics[width=8cm,clip=true]{qmgausscor.eps}  
\includegraphics[width=8cm,clip=true]{dqmgausscor.eps}  
\end{center}  
\caption{\label{fig_cor_gauss}
Same as Fig.~\ref{fig_cor} but the correlation functions 
are evaluated in a random instanton ensemble with Gaussian
fluctuations. }
\end{figure}

 Alternatively, we can determine the correlation function
numerically, using the ``heating'' method. As the name
suggests, this is essentially the inverse of the cooling 
method. We begin from a classical path and determine the 
Gaussian effective potential for small fluctuations around 
the path. For a single instanton, the action is given by
\be 
 S= \int d\tau\, \left( 
 \frac{1}{4} \delta \dot{x}^2(\tau) 
 + 4\eta^2 \left[ 1 -\frac{3}{2\cosh^2(2\eta(\tau-\tau_I))}
 \right]\delta x^2(\tau) \right),
\ee
see Fig.~\ref{fig_veff}. This action has one zero mode
$\delta x(\tau)=-S_0^{-1/2}dx_{cl}(\tau-\tau_I)/(d\tau_I)$
which corresponds to translations of the instanton solution. 
We can eliminate the corresponding non-Gaussian fluctuations
by imposing a constraint on the location of the instanton. 
Using the simple identity
\be
 1 = \int d\tau_I\, \delta( x(\tau_I))\, 
   \left|\dot x(\tau_I)\right|
\ee
we see that the corresponding Jacobian is the velocity
$\dot x(\tau_I)$. We can now perform Monte Carlo calculations 
using the Gaussian action for a multi-instanton configuration. 
The method is illustrated in Fig.~\ref{fig_conf_gauss}. The 
black line shows the classical path and the green path is 
the same path with Gaussian fluctuations included. Clearly, 
this path looks very similar to the full quantum path 
shown in Fig.~\ref{fig_path}. There are still some 
differences, however. We notice, in particular, that the 
fluctuations around the minima of the potential are 
not completely symmetric. This is related to non-Gaussian
effects. We also observe that the heated random instanton
path lacks large excursions from the minima of the potential
that do not lead to a tunneling event. These effects are
due to a combination of instanton interactions and large 
non-Gaussian effects. 

 Correlation functions in the random instanton configurations with 
Gaussian fluctuations included are shown in Fig.~\ref{fig_cor_gauss}. 
We observe that the correlation functions are in much better 
agreement with the exact results than the correlators obtained
from the classical path only. We also see that the correlators
not only describe the splitting between the ground state and
the first excited state but also provide a reasonable 
description of the second excited state. 

\section{Instanton interactions}
\label{sec_int}

\begin{figure}
\begin{center}
\leavevmode
\includegraphics[width=9cm,clip=true]{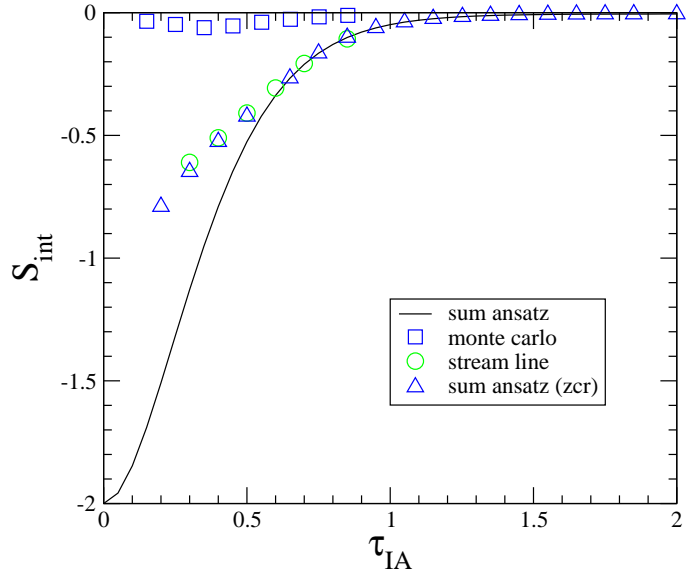}  
\end{center}  
\caption{\label{fig_sint}
Instanton-anti-instanton interaction in units of $S_0$ as
a function of the instanton-anti-instanton separation. The solid
line shows the result in the sum ansatz. The triangles show the
same data plotted as a function of the zero crossing distance.
The streamline interaction is shown as the circles and the 
squares show the effective interaction extracted from the 
cooled intanton-anti-instanton distribution shown in 
Fig.~\ref{fig_zdist}.}
\end{figure}

 Another feature that is missing from the random instanton 
ensemble is the correlation between tunneling events due
to the interaction between instantons. In QCD instanton 
interactions, in particular those mediated by fermions, are
very important and lead to qualitative changes in the instanton
ensemble. In the quantum mechanical model studied here the interaction
between instantons is short range and only leads to relatively small 
effects. These effects can nevertheless be clearly identified in 
very accurate calculations. We refer to \cite{Schafer:1996wv} for
a discussion of the contribution of instanton-anti-instanton 
pairs to the ground state energy. 

 The simplest method for studying the instanton-anti-instanton 
interaction is to construct a trial function and compute 
its action. For the sum ansatz given in equ.~(\ref{sum}) 
the result is shown as the solid line in Fig.~\ref{fig_sint}.
Asymptotically, the action is given by $S_{IA}(\tau_{IA})=2S_0
(1-6\exp(-\eta\tau_{IA})+\ldots)$ where $\tau_{IA}=|\tau_I-\tau_A|$ 
is the instanton-anti-instanton separation. In the opposite limit,
$\tau_{IA}\to 0$, the instanton and anti-instanton annihilate and
the action tends to zero. It is clear, however, that in this limit 
the sum ansatz is at best an approximate solution to the classical 
equation of motion, and it is not obvious how the path should be 
chosen. 

\begin{figure}
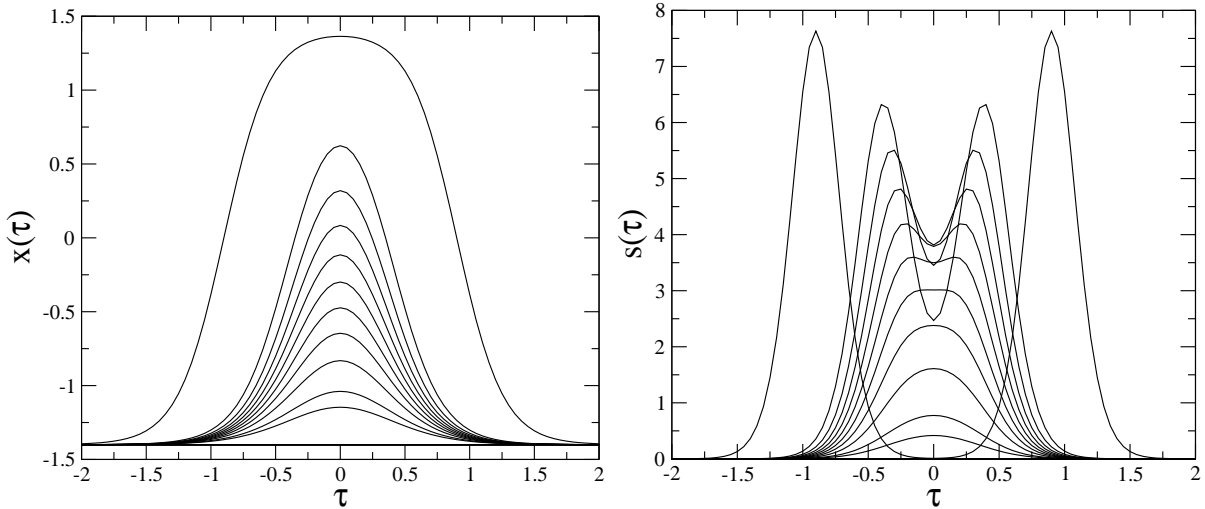

\begin{center}
\leavevmode
\includegraphics[width=8cm,clip=true]{stream_prof.eps}  
\includegraphics[width=7.8cm,clip=true]{stream_act.eps}  
\end{center}  
\caption{\label{fig_stream}
Solution of the streamline equation for an instanton-anti-instanton
pair. Figure a shows the streamline path and the Fig.~b shows the 
action density. The paths correspond to $S/S_0=2.0,1.8,\ldots,0.2,
0.1$.}
\end{figure}

 The best way to deal with this problem is the ``streamline" or 
``valley" method \cite{Balitsky:1986qn,Verbaarschot:1991sq}. The 
method is based on the observation that in the space of all 
instanton-anti-instanton paths there is one almost flat direction 
along which the action slowly varies between $2S_0$ and 0. All other 
directions correspond to perturbative fluctuations. We can force the 
instanton-anti-instanton path to descend along the almost flat 
direction by adding a constraint
\be
S_\xi = \xi(\lambda) \int d\tau\, 
 \left( x(\tau)-x_\lambda(\tau)\right) 
 \frac{dx_\lambda(\tau)}{d\lambda}
\ee
to the classical action. Here, $\lambda$ labels the different
instanton-anti-instanton paths along the streamline and $\xi(\lambda)$
is a Lagrange multiplier. We find the streamline configuration by 
starting from a well separated IA pair and letting the system evolve 
using the method of steepest descent. This means that we have to solve
\be
\label{qm_stream}
\xi(\lambda)\frac{dx_\lambda(\tau)}{d\lambda} = 
  \left.\frac{\delta S}{\delta x (\tau)}\right|_{x=x_\lambda},
\ee
with the boundary condition that $x_{\lambda=0}(\tau)\simeq x_{sum}(\tau)$ 
corresponds to a well separated instanton-anti-instanton pair. Note that
$\xi(\lambda)$ is an arbitrary function that reflects the reparametrization 
invariance of the streamline solution. A sequence of paths obtained by 
solving equ.~(\ref{qm_stream}) numerically is shown in Fig.~\ref{fig_stream}. 
We also show the action density $s=\dot x^2/4+V(x)$. We can see clearly how 
the two localized solutions merge and eventually disappear as the 
configuration progresses down the valley.

 There is no unique way to parametrize the streamline path 
and extract the instanton-anti-instanton action as a function of the 
separation between the tunneling events. The simplest possibility
is to used the distance between the zero crossings $\tau_z$. This 
definition has the advantage of being very easy to use, but it 
prevents us from exploring the part of the streamline trajectory
where the instanton and anti-instanton are so close that the
path never crosses zero. In Fig.~\ref{fig_sint} we compare results 
for $S_{IA}(\tau_z)$ obtained from the sum ansatz and the streamline 
solution. We observe that for instanton separations $\tau_z>0.3$ the 
results are very similar. We also note, however, that for $\tau_z
<0.6$ the zero crossing distance is quite different from the
parameter $\tau_I-\tau_A$ that appears in the sum ansatz. 

\begin{figure}
\begin{center}
\leavevmode
\includegraphics[width=9cm,clip=true]{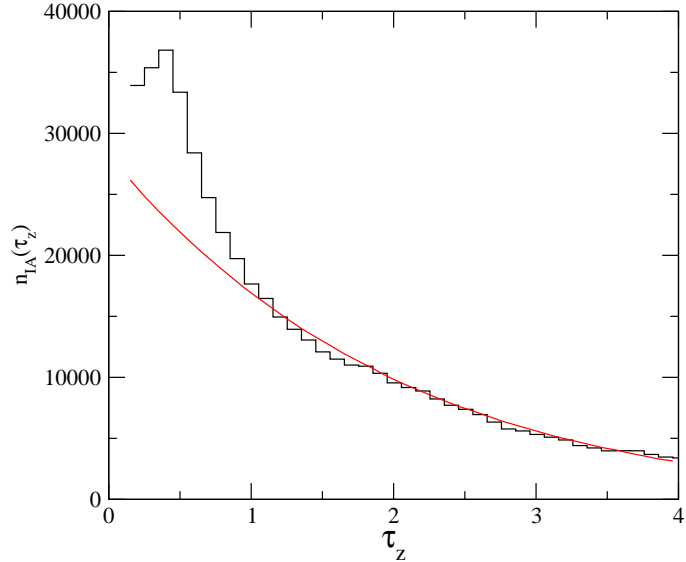}  
\end{center}  
\caption{\label{fig_zdist}
Distribution of instanton-anti-instanton separations after 
10 cooling sweeps. The separation was extracted from the 
distance of the zero-crossings in the cooled configuration.}
\end{figure}

 One can show that the ambiguities that arise in trying to
define the instanton-anti-instanton interaction at short 
distance correspond to similar ambiguities that arise in the 
perturbative expansion because and are related to the factorial 
growth of higher order terms in the expansion. Only the sum of 
the perturbative and the instanton contribution is well defined 
and leads to unique predictions for the groundstate energy. 
In these lectures we shall not discuss this problem any 
further. Instead, we will study the question whether the
full quantum configurations contain evidence of the correlations 
between instantons that the classical instanton-anti-instanton
interaction implies. 

\begin{figure}
\begin{center}
\leavevmode
\includegraphics[width=9cm,clip=true]{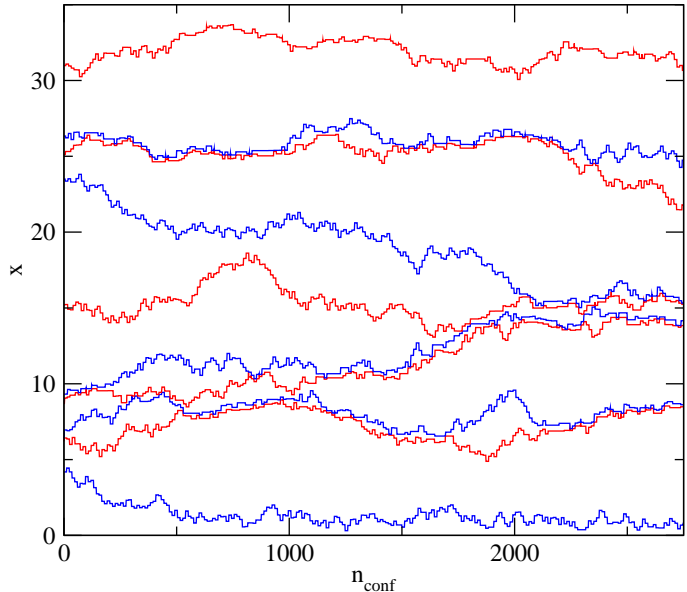}    
\end{center}  
\caption{\label{fig_iconf}
Typical instanton configuration in an interacting instanton 
calculation. The figure shows the location $x$ of the 
first 10 instantons (blue) and anti-instantons (red) 
over a period of 3000 configurations.}
\end{figure}

 In Fig.~\ref{fig_zdist} we show a histogram of the 
instanton-anti-instanton separation determined in Monte 
Carlo simulations of the quantum mechanical partition function. 
The data were obtained by measuring the zero crossing distance
after 10 cooling sweeps. For comparison we also show the $IA$ 
distribution in the random instanton gas. We observe that there 
is an enhancement of close $IA$ pairs which corresponds to an 
attractive instanton-anti-instanton interaction. The exact magnitude 
of this enhancement depends sensitively on the number of cooling 
sweeps. As emphasized in the previous paragraph, this is not necessarily 
a shortcoming of the cooling method. We can try to translate the 
enhancement in the $IA$ distribution into an effective interaction 
using the classical relation $n(\tau_{IA})\sim n_0(\tau_{IA})\exp(-
S_{IA}(\tau_{IA}))$. Here, $n(\tau_{IA})$ is the $IA$ distribution, 
$n_0(\tau_{IA})$ is the distribution in the random theory and $S_{IA}
(\tau_{IA})$ is the instanton-anti-instanton interaction. The result 
is also shown in Fig.~\ref{fig_sint}. We observe that the interaction 
extracted from the $IA$ distribution is significantly weaker than the 
classical result. This may imply that the full quantum interaction is 
weaker than the classical result, or that too many close pairs are 
lost during cooling. This question can be studied in more detail using 
the methods discussed in Sect.~\ref{sec_dens}. 

 Finally, we address the question how to include correlations 
between tunneling events in the instanton calculation. For 
this purpose we include the instanton-anti-instanton interaction
in the instanton liquid partition function equ.~(\ref{rilm}). 
In this context we again have to address the problem of close
instanton-anti-instanton pairs. The simplest approach is to
add a short range repulsive core which excludes configurations
that are not semi-classical. The hard core interaction can be 
adjusted in order to reproduce the $IA$ distribution found in the 
cooling calculation. In practice we have chosen $S_{core}(\tau_{IA})=
A_c \exp(-\tau_{IA}/\tau_c)$ with $A_c=3$ and $\tau_c=0.3$. In 
Fig.~\ref{fig_iconf} we show a typical set of instanton and 
anti-instanton trajectories from an interacting instanton calculation. 
Correlations between instantons are clearly visible. In particular, 
we observe that a number of close instanton-anti-instanton pairs 
are formed. We have also studied correlation functions in the 
interacting instanton ensemble. We find that differences as compared 
to the random ensemble are rather subtle and a detailed study of
non-Gaussian effects is necessary in order to establish the 
importance of instanton interactions.

\section{Summary}
\label{sec_sum}

  In these lectures we presented Monte Carlo methods for 
studying the euclidean path integral in Quantum Mechanics.
We also supply a set of computer codes written in fortran 
that were used to generate the data shown in the figures.
We encourage the reader to play around with these programs
in order to get a deeper appreciation of the path integral 
and of Monte Carlo methods. 

 We should note that Monte Carlo calculations of the 
euclidean path integral are an extremely poor way to compute 
the spectrum or the correlation functions of the anharmonic
oscillator. The code based on diagonalizing the Hamiltonian
is both much faster and much more accurate than the Monte
Carlo codes. The purpose of the Monte Carlo codes is 
entirely pedagogical. However, if we proceed from quantum 
mechanics to systems involving many more degrees of freedom, 
such as four-dimensional field theories, Hamiltonian methods
become more and more impractical and Monte Carlo calculations
based on the euclidean path integral provide the most efficient 
method for computing the spectrum and the correlation functions
known to date. 

 We also discussed Monte Carlo methods for studying the 
contribution of instantons to the euclidean path integral. 
In the case of the double well potential there is a parameter,
$\eta$, which controls the instanton action $S_0=4\eta^3/3$.
If $S_0\gg 1$ then instantons are easily identified but the
tunneling rate is small. If $S_0\sim 1$ then instantons are 
very abundant but it is hard to determine the instanton density  
precisely. We focused on the case $S_0\sim 3$ which is at 
the boundary of the semi-classical regime. Even though the 
expansion parameter $1/S_0$ is not very small the instanton
density is still well determined and agrees with the level
splitting. We also noticed, however, that non-Gaussian 
effects are important in this regime. 

 Ultimately, we are interested in the question to what 
extent these results can be generalized to QCD. In QCD 
there is no free parameter that controls the applicability 
of the semi-classical expansion. Unlike the case of the 
double well potential, instantons in QCD can have any 
size. Asymptotic freedom implies that the action of small
instantons is big, but the action of instantons with size
$\rho\sim\Lambda_{QCD}^{-1}$ is of order one. Nevertheless, 
lattice calculations support the idea that the tunneling density 
is sizable, $(N/V)\simeq \Lambda_{QCD}^4 \simeq 1\,{\rm fm}^{-4}$, 
and that instantons do not strongly overlap \cite{Chu:1994vi}. 
The typical instanton size is found to be $\rho\sim (0.3-0.4)$ 
fm which implies a typical instanton action $S_0\simeq 
(5-10)$. The reason that the density is big even though 
the action is significantly larger than one is related 
to the fact that QCD instantons have many more collective 
coordinates, 12 (4 coordinates, 1 size, 7 color angles) 
compared to just one in the case of the double well potential.
As a consequence the pre-exponential factor in the tunneling 
rate is numerically large. 

 There are some important differences between QCD and the
double well potential. In a typical lattice QCD configuration 
quantum fluctuations of the gauge field are much bigger than 
the classical gauge fields associated with instantons. This 
implies that one cannot ``see'' instantons in the gauge 
configurations in the same way that one can immediately 
identify tunneling events in the quantum mechanical paths.
Compare, for example, Fig.~\ref{fig_path} with Fig.~1
in \cite{Chu:1994vi}. Only after some amount of cooling do 
instantons emerge from the quantum noise. On the other hand, 
fermions provide an important diagnostic tool in QCD that is 
not available in the simple bosonic model analyzed in these 
lectures. Instantons lead to localized chiral zero modes of 
the Dirac operators that can easily be identified even in noisy
quantum configurations.

\newpage 

\appendix
\section{Computer Codes}

\noindent 
All programs are written in standard fortran 77, have extensive 
comments and do not require any libraries. Some of the programs 
contain subroutine for generating random numbers or for diagonalizing 
matrices that were taken from Numerical Recipes (Numerical Recipes
in Fortran, W.~H.~Press, S.~A.~Teukolsky, W.~T.~Vetterling and
B.~D.~Flannery, Cambridge University Press). 
\vspace*{0.5cm}

\noindent
{\bf 1. qmdiag.for}

\noindent 
This programs computes the spectrum and the eigenfunctions
of the anharmonic oscillator. The results are used in order 
to compute euclidean correlation functions. 

\noindent
Input: {\tt fort.05}

\begin{tabular}{p{2.2cm}p{14cm}}
$f$ & minimum of anharmonic oscillator potential $V(x)=(x^2-f^2)^2$ \\
$N$ & dimension of basis used for diagonalizing $H$ (choose $N\geq 40$)\\
$\omega_0$ & unperturbed oscillator frequency (choose $\omega_0\sim 4f$) 
\end{tabular}
\vspace*{0.3cm}

\noindent
Output: {\tt qmdiag.dat}

\begin{tabular}{p{2.2cm}p{13cm}}
$E_n$   &  eigenvalue of Hamiltonian \\
$c_n$   &  dipole matrix element $c_n^2=|\langle 0|x|n\rangle|^2$\\
$d_n$   &  quadrupole matrix element $d_n^2=|\langle 0|x^2|n\rangle|^2$\\
$e_n$   &  quadrupole matrix element $d_n^2=|\langle 0|x^3|n\rangle|^2$\\
$\psi(x)$ & ground state wave function \\
$\langle x(0)x(\tau)\rangle$ & euclidean correlation function, 
           also for $x^2$ and $x^3$\\
$d\log\Pi/(d\tau)$ & log derivative of $\Pi(\tau)= \langle x(0)x(\tau)
           \rangle$ \\
$Z(\beta)$ & partition function  
\end{tabular}

\newpage
\noindent
{\bf 2. qm.for}

\noindent 
This programs computes correlation functions of the anharmonic
oscillator using Monte Carlo simulations on a euclidean 
lattice. 

\noindent
Input: {\tt fort.05}

\begin{tabular}{p{2.0cm}p{13cm}}
$f$ & minimum of anharmonic oscillator potential  $V(x)=(x^2-f^2)^2$  \\
$n$ & number of lattice points in the euclidean time direction 
       ($n\sim 800$) \\
$a$ & lattice spacing ($a\sim 0.05$)  \\
$ih$& $ih=0$ cold start $x_i=-f$, $ih=1$ hot start $x_i=ran()$   \\
$n_{eq}$ & number of equilibration sweeps before first measurement
       $(n_{eq}\sim 100)$  \\
$n_{mc}$ & number of Monte Carlo sweeps ($n_{mc}\sim 10^5$)  \\
$\delta x$ &width of Gaussian distribution used for Monte Carlo
            update $x_i^{(n)}\to x_i^{(n+1)}$ ($\delta x \sim 0.5$) \\ 
$n_{p}$ & number of points on which correlation functions are 
          measured $\langle x_ix_{i+1}\rangle,\ldots,$
          $\langle x_i x_{i+n_p}\rangle$ $(n_p\sim 20)$  \\
$n_{mea}$ & number of measurements of the correlation function
            in a given Monte Carlo configuration ${x_i}$ ($n_{mea}\sim 5$)  \\
$n_{pri}$ & number of Monte Carlo configurations between output 
           of averages to output  file ($n_{pri}\sim 100$)   
\end{tabular}
\vspace*{0.3cm}

\noindent
Output: {\tt qm.dat}

\begin{tabular}{p{2.0cm}p{13cm}}
$S_{tot}$ &average total action per configuration  \\
$V_{av},T_{av}$ &  average potential and kinetic energy \\
$\langle x^n\rangle$  & expectation value $\langle x^n\rangle$ 
             ($n=1,\ldots,4$)      \\
$\Pi(\tau)$ & euclidean correlation function $\Pi(\tau) = 
              \langle O(0)O(\tau)\rangle$ for $O=x$, $x^2$, $x^3$.
              Results are given in the format $\tau, \Pi(\tau)$,
              $\Delta\Pi(\tau)$, $d\log(\Pi)/(d\tau)$, $\Delta[ 
              d\log(\Pi)/(d\tau)]$, where $\Delta\Pi(\tau)$ is 
              the statistical error in $\Pi(\tau)$.  
\end{tabular}

\newpage
\noindent
{\bf 3. qmswitch.for}

\noindent
The program {\tt qmswicth.for} computes the free energy $F=-T\log(Z)$ 
of the anharmonic oscillator using the method of adiabatic switching 
between the harmonic and the anharmonic oscillator. The action is
$S_\alpha=S_0+\alpha(S-S_0)$. The code switches from $\alpha=0$ to 
$\alpha=1$ and then back to $\alpha=0$. Hysteresis effects are
used in order to estimate errors from incomplete equilibration.
Most input parameters are the same as in {\tt qm.for}. Additional 
parameters are given below. 

\noindent
Input: {\tt fort.05}

\begin{tabular}{p{2.0cm}p{13cm}}
$\omega_0$   & oscillator constant of the reference system
                ($\omega_0\sim 4f$) \\
$n_{switch}$ & number of steps in adiabatic switching ($n_{switch}
                \sim 20$)  \\
\end{tabular}
\vspace*{0.3cm}

\noindent 
Output: {\tt qmswitch.dat}

\noindent
The output file contains many details of the adiabatic switching 
procedure. The final result for the free energy is given as $F=F_0
+\delta F$, where $F_0$ is the free energy of the harmonic oscillator 
and $\delta F$ is the integral over $\alpha$. We estimate the uncertainty 
in the final result as $F\pm \Delta F(stat)\pm \Delta F(equ) \pm \Delta 
F(disc)$, where $\Delta F(stat)$ is the statistical error, $\Delta F(equ)$ 
is due to incomplete equilibration (hysteresis), and $\Delta F(disc)$ is 
due to discretizing the $\alpha$ integral.

\vspace*{0.3cm}

\noindent
{\bf 4. qmcool.for}

\noindent 
This programs is identical to ${\tt qm.for}$ except that 
expectation values are measured both in the original 
and in cooled configurations. We only specify additional 
input parameters. 

\noindent
Input: {\tt fort.05}

\begin{tabular}{p{2.0cm}p{13cm}}
$n_{st}$   & number of Monte Carlo configurations between successive 
             cooled configurations. The number of cooled configurations 
             is $n_{conf}/n_{st}$ ($n_{st}\sim 20$). \\
$n_{cool}$ & number of cooling sweeps ($n_{cool}\sim 50$) 
\end{tabular}
\vspace*{0.3cm}

\noindent
Output: {\tt qmcool.dat}

\begin{tabular}{p{2.0cm}p{13cm}}
$\Pi(\tau)$ & correlation functions are given in the same format 
               as in ${\tt qm.dat}$ \\
$N_{I+A}$   & total number of instantons extracted from number 
               of zero crossings \\
            & as a function of the number of cooling sweeps \\
$S_{tot}$   & total action vs number of cooling sweeps  \\
$S/N$       & action per instanton. $S_0$ is the continuum 
               result for one instanton 
\end{tabular}
\vspace*{0.3cm}

\newpage
\noindent 
{\bf 5. qmidens.for}

\noindent 
The program {\tt qmidens.for} calculates non-Gaussian corrections
to the instanton density using adiabatic switching between the 
Gaussian action and the full action. The calculation is performed
in both the zero and one-instanton-sector. The details of the
adiabatic switching procedure are very similar to the method
used in {\tt qmswitch.for}. Note that the total length of 
the euclidean time domain, $\beta=na$, cannot be chosen too
large in order to suppress transitions between the one-instanton
sector and the three, five, etc.~instanton sector. Most input 
parameters are defined as in {\tt qm.for}. 

\noindent
Input: {\tt fort.05}

\begin{tabular}{p{2.0cm}p{13cm}}
$n_{switch}$ & number of steps in adiabatic switching ($n_{switch}
                \sim 20$)  \\
\end{tabular}
\vspace*{0.3cm}

\noindent 
Output: {\tt qmidens.dat}

\noindent
The output file contains many details of the adiabatic 
switching procedure. The final result for the instanton
density is compared to the Gaussian (one-loop) approximation.
Note that the method breaks down if $f$ is too small or 
$\beta$ is too large. 

\newpage
\noindent
{\bf 6. rilm.for}

\noindent 
This program computes correlation functions of the anharmonic
oscillator using a random ensemble of instantons. The 
multi-instanton configuration is constructed using the 
sum ansatz. Note that, in contrast to RILM calculations 
in QCD, the fields and correlation functions are computed
on a lattice.

\noindent
Input: {\tt fort.05}

\begin{tabular}{p{2.0cm}p{13cm}}
$f$ & minimum of anharmonic oscillator potential $V(x)=(x^2-f^2)^2$  \\
$n$ & number of lattice points in the euclidean time direction 
       ($n\sim 800$) \\
$a$ & lattice spacing ($a\sim 0 05$)  \\
$N_{I+A}$ & number of instantons (has to be even).  The program 
       displays the one and two-loop result for the parameters 
       $(f,\beta=na)$. \\
$n_{mc}$ & number of configurations ($n_{mc}\sim 10^3$) \\
$n_{p}$ & number of points on which correlation functions are 
          measured $\langle x_ix_{i+1}\rangle,\ldots,$
          $\langle x_i x_{i+n_p}\rangle$ $(n_p\sim 20)$ \\
$n_{mea}$ & number of measurements of the correlation function
            in a given Monte Carlo configuration ${x_i}$ 
            ($n_{mea}\sim 5$) \\
$n_{pri}$ & number of Monte Carlo configurations between output 
            of averages to output  file ($n_{pri}\sim 100$)
\end{tabular}
\vspace*{0.3cm}

\noindent
Output: {\tt rilm.dat}

\begin{tabular}{p{2.0cm}p{13cm}}
$S_{tot}$ &average total action per configuration. \\
$V_{av},T_{av}$ &  average potential and kinetic energy.\\
$\langle x^n\rangle$  & expectation value $\langle x^n\rangle$ 
            ($n=1,\ldots,4$)     \\
$\Pi(\tau)$ & euclidean correlation function $\Pi(\tau) = 
            \langle O(0)O(\tau)\rangle$ for $O=x$, $x^2$, $x^3$.
            Results are given in the format $\tau$, $\Pi(\tau)$,
            $\Delta\Pi(\tau)$, $d\log(\Pi)/(d\tau)$, $\Delta[ 
            d\log(\Pi)/(d\tau)]$.
\end{tabular}
\vspace*{0.3cm}

\newpage
\noindent 
{\bf 7. rilm\_gauss.for}

\noindent 
This program generates the same random instanton ensemble as 
{\tt rilm.for} but it also includes Gaussian fluctuations 
around the classical path. This is done by performing a
few heating sweeps in the Gaussian effective potential.
Most input parameters are defined as in {\tt rilm.for}. 
Additional input parameters are given below. 

\noindent
Input: {\tt fort.05}

\begin{tabular}{p{2.0cm}p{13cm}}
$n_{heat}$ & number of heating steps ($n_{heat} \sim 10$) \\
$\delta x$ & coordinate update ($\delta x\sim 0.5$) 
\end{tabular}

\vspace*{0.5cm}

\noindent 
{\bf 8. iilm.for}

\noindent 
This program computes correlation functions of the anharmonic
oscillator using an interacting ensemble of instantons. The 
multi-instanton configuration is constructed using the sum 
ansatz. The configuration is discretized on a lattice and
the total action is computed using the discretized lattice 
action. Very close instanton-anti-instanton pairs are 
excluded by adding an nearest neighbor interaction with 
a repulsive core. Most input parameters are the same as
in ${\tt rilm.for}$. Additional input parameters are 

\noindent
Input: {\tt fort.05}

\begin{tabular}{p{2.0cm}p{13cm}}
$\tau_{core}$ & range of hard core interaction ($\tau_{core}\sim 0.3$) \\
$A_{core}$    & strength of hard core interaction ($A_{core}\sim 3.0$) \\
$dz$          & average position update ($dz\sim 1$)
\end{tabular}

\end{document}